\documentclass[aps,prl,twocolumn,superscriptaddress,showpacs,amssymb,amsmath]{revtex4}
\usepackage{graphicx}
\usepackage{txfonts}
\usepackage{color}
\graphicspath{{./figures/}}
\DeclareGraphicsExtensions{.eps}
\newcommand{\figwidth}{\linewidth}

\newcommand{\csic}{Instituto de Estructura de la Materia, CSIC,
Serrano 123, 28006 Madrid, Spain}
\newcommand{\arizona}{Physics Department, University of Arizona\\
1118 E.\ Fourth Street, P.O. Box 210081,
Tucson, AZ 85721, USA}
\newcommand{\geneve}{D\'epartement de Physique Th\'eorique, Universit\'e de Gen\`eve, 
CH-1211 Gen\`eve 4, Switzerland}
\newcommand{\ipcms}{Institut de Physique et Chimie des Mat{\'e}riaux de 
Strasbourg, UMR 7504, CNRS-UdS,\\
23 rue du Loess, BP 43, 67034 Strasbourg Cedex 2, France}

\begin{document}

\title{Scattering phase of quantum dots: Emergence of universal behavior}

\author{Rafael A.\ Molina}
\affiliation{\csic}
\author{Rodolfo A.\ Jalabert}
\affiliation{\ipcms}
\author{Dietmar Weinmann}
\affiliation{\ipcms}
\author{Philippe Jacquod}
\affiliation{\arizona}
\affiliation{\geneve}

\date{\today}

\begin{abstract}

We investigate scattering through chaotic ballistic quantum dots in the Coulomb 
blockade regime. Focusing on the scattering phase, we show that large universal 
sequences emerge in the short wavelength limit, where phase lapses of $\pi$ 
systematically occur between two consecutive resonances. Our results are corroborated 
by numerics and are in qualitative agreement with existing experiments.
\end{abstract}



\maketitle

Quantum mechanics fundamentally differs from classical mechanics in that time 
evolutions are determined by complex probability amplitudes instead of real 
probabilities. The associated phase is a key element to understand mesoscopic 
transport experiments on Aharonov-Bohm (AB) conductance oscillations, weak 
localization, and conductance fluctuations \cite{imry}.
However, these transport measurements, like any other measurement, do not directly 
measure scattering phases. 
In their phase-sensitive experiment, Yacoby \textit{et al.} pioneered an entirely 
new field of mesoscopic physics, by embedding a ballistic quantum dot (QD) in one 
arm of an AB interferometer \cite{jacoby95}. The sustained interest in these and 
following~\cite{schuster97,sigrist04,avinum05} experiments, that persists until 
today, relies on the difficulty to consistently and generically explain
these measurements. 

\begin{figure}
\centerline{\includegraphics[width=\figwidth]{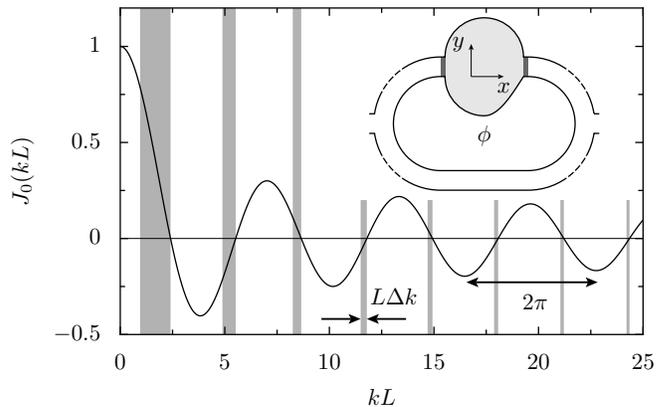}}
\caption{\label{fig:bessel} 0$^{\rm th}$ order Bessel function $J_0(kL)$. The shaded 
areas indicate regions where $\bar{D}_n$ is negative. For two-dimensional systems, 
the width of these intervals, and the probability of not obtaining a phase lapse 
between two resonances, decrease as $L \Delta k \sim (kL)^{-1}$. 
Inset: 
Aharonov-Bohm interferometer, threaded by a flux $\phi$, with an asymmetric 
dot (shaded) tunnel-embedded in its upper arm. The distance between the entrance 
and exit points of the dot is $L$. The dashed lines on the arms of the 
interferometer stand for any number of potential additional leads.}
\end{figure}

When a QD in the Coulomb-blockade (CB) regime is placed in one arm of a mesoscopic 
ring threaded by a flux $\phi$ (see the inset of Fig.~\ref{fig:bessel}) the 
conductance through the ring reads \cite{levy95,aharony2002}
\begin{equation}
g=g_0+\sum_{n} g_n \cos{(2 \pi n \phi/\phi_0+\beta_n)}\, ,
\end{equation}
with the quantum of flux $\phi_0$.
Measuring the lowest harmonics of the AB oscillations in $\phi$ allows to extract the 
phase $\beta_1$ as a function of a gate voltage $V_\mathrm{g}$ applied to the QD. 
Under suitable conditions $\beta_1$ can be related to the scattering phase of the 
QD \cite{aharony2002}. A multi-terminal configuration is required in order to get a 
one-to-one relationship between the two phases, while in a two-terminal set-up 
$\beta_1$ can only take the values $0$ or $\pi$ \cite{levy95,schuster97}. 
In the multi-terminal case a gradual increase of $\pi$ in the transmission
phase is obtained as a function of the gate voltage for every CB peak, in
agreement with the Friedel sum rule \cite{lee99,taniguchi99,levy00}. 
Abrupt lapses of $\pi$ occur between resonances in both configurations at 
values of $V_\mathrm{g}$ for which the transmission amplitude of the QD is so small 
that the AB oscillations are below the
experimental visibility threshold. For relatively large QD with a few hundred 
electrons \cite{jacoby95,schuster97}, the lapses were seen to appear systematically 
between every consecutive pair of resonances. 
This surprising behavior, termed universal, is observed in two- or multi-terminal 
configurations. 
Smaller dots were more 
recently investigated, where the number of electrons was tuned from about twenty down 
to zero \cite{avinum05}. A crossover from the universal regime, where
successive peaks are in phase, to a mesoscopic regime, where phase lapses occur in a 
random fashion, was observed when decreasing the number of electrons in the QD by the 
action of $V_\mathrm{g}$. Multi-terminal configurations with one QD in each arm of 
the AB interferometer have also been investigated \cite{sigrist04}, which determined 
the important role of the magnetic field in the phase lapses. 
The understanding of the crossover from the mesoscopic to universal regime and the 
role of the symmetries are the main goals of our work.

For lateral QD in the CB regime a single lead channel dominantly couples to the QD,
and transport through the QD is characterized by a $2\times2$ scattering matrix
\begin{equation}
\begin{aligned}
S = \left(\begin{array}{cc}
r & t' \\
t & r'
\end{array}\right)
 = e^{i \alpha} \left(\begin{array}{cc}
i e^{i\xi} \cos{\theta} & e^{i\eta}\sin{\theta} \\
e^{-i\eta}\sin{\theta} & ie^{-i\xi}\cos{\theta}
\end{array}\right)
\, .
\end{aligned}
\label{eq:sm}
\end{equation}
We note $t$ ($t'$) and $r$ ($r'$) the transmission and reflection amplitudes, 
respectively, for particles coming from the left (right) of the QD. The angle 
$\theta$ is restricted to the interval $[0,\pi)$, while the phases $\alpha$, $\eta$ 
and $\xi$ are defined on  $[0,2\pi)$. The scattering phase $\alpha$ is related to the 
density of states of the QD through the Friedel sum rule 
\cite{levy95,lee99,taniguchi99,levy00}. When considering the phase evolution as a function 
of an external parameter (like $V_\mathrm{g}$) it is convenient to work with the 
accumulated phase $\alpha_{\rm c}$, whose range of definition is not restricted to 
the interval $[0,2\pi)$. 

Eq.~(\ref{eq:sm}) represents the most general $2\times2$ unitary matrix. 
When time-reversal invariance is present, one has $\eta=0$ or $\pi$.
Right-left parity symmetry would restrict $\xi$ to either $0$ or $\pi$. 
Here, we consider generic QDs with arbitrary $\xi$. 
When the many-body problem is considered 
in its full complexity, $S$ represents an effective one-particle scattering 
matrix that can be obtained for instance through the embedding method 
\cite{molina03,molina_unp}. If we restrict the description of the many-body 
problem to the constant-interaction model (CIM) the CB phenomena can be 
interpreted in terms of single-particle quantities \cite{alhassid}, and
close to resonances, $S$ is given by the transmission and reflection amplitudes of 
single-particle states in a mean-field potential. When the underlying classical 
scattering is chaotic, $S$ has well defined statistical properties determined by 
the symmetries of the problem only \cite{JP}. In structures with time-reversal symmetry, $t= e^{i \alpha}\sin{\theta}$ and, even if $t$ is a complex continuous function of the real variable $V_\mathrm{g}$, the phase $\alpha$ exhibits a phase lapse of $\pi$ whenever $t$ vanishes. 

Within the CIM and in the absence of a magnetic field the one-particle
wave-functions can be chosen to be real. With the exception of peculiar 
cases with very different values of the resonance widths between consecutive 
resonances, a zero of the transmission generically appears between two 
successive resonances depending on the sign of \cite{levy00}
\begin{equation}
D_n=\gamma_n^{\mathrm{l}}\gamma_n^{\mathrm{r}}\gamma_{n+1}^{\mathrm{l}}\gamma_{n+1}^{\mathrm{r}} \ .
\end{equation}
The partial-width amplitude (or effective coupling) of the eigenstate
$\psi_n$ of the QD is given by \cite{alhassid}
\begin{equation}
\gamma_n^{\mathrm{l}(\mathrm{r})} = \left(\frac{\hbar^2kP_\mathrm{c}}{m}\right)^{1/2}
\int_{0}^{W} {\rm d}y\, \Phi_{\mathrm{l}(\mathrm{r})}(y) \, \psi_n(x^{\mathrm{l}(\mathrm{r})},y) \ .
\end{equation}
Here, $\Phi_{\mathrm{l}(\mathrm{r})}$ is the first transversal subband
wave-function in the the left (right) lead of width $W$, the integration is along 
the transverse coordinate $y$ at the entrance or exit of the QD located at 
$x=x^{\mathrm{l}(\mathrm{r})}$, $P_\mathrm{c}$ is the transparency of the tunnel 
barriers, $k$ is the Fermi wave-vector in the leads, and $m$ is the electron mass. 
In lattice models with one-dimensional leads \cite{levy00,aharony2002} the two 
partial-width amplitudes are simply proportional to the value  of the wave-function 
at the extreme point connecting the QD to the corresponding lead. 

We call $\gamma_n^{\mathrm{l}}\gamma_n^{\mathrm{r}}$ the {\it parity}
of the $n^{\rm th}$ resonance\cite{levy00}. When $D_n>0$ (equal parity of 
the $n^{\rm th}$ and $(n+1)^{\rm st}$ resonances) there is one zero (or an odd 
number of zeros) between the $n^{\rm th}$ and $(n+1)^{\rm st}$ resonances, while for 
$D_n<0$ (opposite parities) there is no zero (or an even number of zeros) between the 
two resonances. For one-dimensional scatterers, $D_n$ is always negative, which 
results in the absence of transmission zeros reflecting the impossibility of 
obtaining destructive interfering paths in one dimension\cite{lee99}.
Numerical simulations on two-dimensional disordered lattice systems, on the other 
hand yield an equal probability for positive or negative $D_n$ \cite{levy00}, 
which is at odds with the experimental observation of a universal regime with
zeros between any two consecutive resonances. Several theoretical refinements
have been proposed to solve the puzzle
\cite{levy00,hacken,gefen99,silvestrov00,golosov06,bertoni07,vonDelft07,goldstein07,bergfield10},
where specific geometries or effective couplings, or extensions of the 
CIM are considered.

The latter path is in principle the most natural one, and was actually already suggested
in Ref.~\cite{jacoby95}. However, the small
sizes that can be handled within a full many-body description make the universal 
regime hardly reachable. Moreover, even if the influence of electron-electron
interaction can be spotted in some circumstances 
\cite{vonDelft07,molina_unp,bergfield10}, the corresponding results are not generic. 
On the other hand, the CIM gives an excellent description of the statistical 
distribution of the height of the CB peaks \cite{jala92,alhassid}, which is 
determined by the one-particle widths $\gamma^\mathrm{l,r}_n$. It is then expected 
that the statistics of the resonance parities and consequently of the transmission 
zeros are within the reach of the CIM. Below, assuming that this is the case, we 
incorporate well known correlations of quantum chaotic wavefunctions in our analysis 
of $D_n$. In this way we
account for the transition from a mesoscopic to a universal regime. 

Long-range wavefunction correlations in quantum chaotic systems were first pointed 
out by Berry \cite{berry77}, who suggested to model the wavefunctions as random 
superpositions of plane waves with fixed energy $\hbar^2k^2/(2m)$. For a 
two-dimensional chaotic billiard with eigenfunctions $\psi_n$, 
this gives the wavefunction correlator
\begin{equation}
\overline{\psi_n({\bf r})\psi_n({\bf r'})} = 
J_0(k|{\bf r}-{\bf r'}|)/{\cal A}  \ .
\label{BerryBessel}
\end{equation}
The bar stands for a local average, ${\cal A}$ is the area of the billiard and 
$J_0$ is the $0^{\rm th}$ Bessel function of the first kind.
Corrections to Eq.~(\ref{BerryBessel}) appear in confined systems when the 
observation points approach the boundaries \cite{urbina04}, 
which is the case we are interested in. For ${\bf r} \ne {\bf r'}$, these 
corrections are however small and we will neglect them. 
Eq.~(\ref{BerryBessel}) has been successfully used to explain the long-range (in energy) modulation 
of the peak-height distribution in the CB regime \cite{narimanov}. 

If we call $L$ the distance from the entrance to the exit of the QD and
assume that successive eigenfunctions are uncorrelated we have for the case 
$W \ll L$ of relevance for us
\begin{eqnarray}
{\bar D_n} \sim J_0(k_nL) J_0(k_{n+1}L) \ .
\label{eq:besbes}
\end{eqnarray}
Taking $L$ as the typical linear dimension of the QD, we have 
$k_{n+1}-k_n \simeq \Delta k=\pi/(kL^2)$. Thus, ${\bar  D}_n$ is negative when the 
Bessel functions in Eq.~(\ref{eq:besbes}) have different signs, which only happens if 
$kL$ falls in an interval of length $L \Delta k$ before one zero of $J_0(kL)$. 
This is sketched in Fig.~\ref{fig:bessel}. In the semiclassical limit $kL \gg 1$ 
we have $J_0(kL) \sim \sqrt{2/(\pi k L)} \cos{\left(kL-\pi/4\right)}$ and 
therefore the probability ${\cal P}$ of obtaining a negative ${\bar  D}_n$  can be
estimated as the ratio of $2 L \Delta k$ over the period $2\pi$, that is,
\begin{equation}
{\cal P} \sim \frac{1}{kL} \ .
\label{eq:prob}
\end{equation}
This simple analysis explains why when the dot is progressively filled the
equal-parity case (${\bar  D}_n>0$) results with probability 
approaching one. Moreover,
it also shows that the sequence of peaks and zeros appears over intervals 
in $kL$ that are of length $\pi$, containing a number $~kL$ of resonances.
The sole assumption that wave-functions have quantum chaotic correlations thus 
predicts the emergence of large universal sequences of resonances and transmission 
zeros. This is our main result.

\begin{figure}
\centerline{\includegraphics[width=\figwidth]{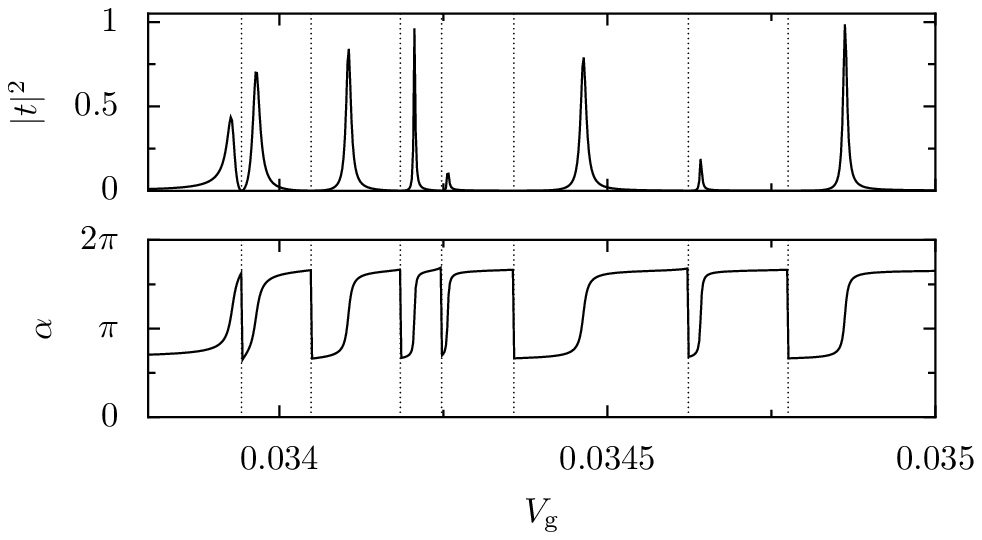}}
\caption{\label{fig:condandphase} Transmission $|t|^2$ and scattering
phase $\alpha$ as a function of the applied gate voltage $V_\mathrm{g}$. 
The number of electrons in the dot varies between 174 and 182. Smooth 
increases of $\pi$ in $\alpha$ obtained for each resonance alternate with 
phase lapses of $-\pi$ when the transmission vanishes (dotted vertical lines).}
\label{fig:condandpha}
\end{figure}

In order to illustrate and test our analytical prediction we have performed
numerical calculations of the transmission amplitude of spinless electrons
through a noninteracting QD 
with the shape of a desymmetrized stadium billiard, where one of the quarter 
circles is replaced by a cosine curve. This is sketched in the inset of 
Fig.~\ref{fig:bessel}. The QD is connected to leads through tunnel barriers 
as in Ref.~\cite{jala92}. The parameter of variation
is the gate voltage $V_{\rm g}$ which induces changes in the wave-vector $k$ 
within the QD. Phase increases of $\pi$ are correlated with transmission
peaks, while phase lapses of $\pi$ are assigned at the values of $V_{\rm g}$ 
where the transmission vanishes.
We show in Fig.~\ref{fig:condandpha} a sequence of eight resonances for
174 to 182 electrons on the dot, similar to the experiments of Refs.~\cite{jacoby95,schuster97}. 
This sequence exhibits a perfect alternation of resonances and zeros. We found that this is the generic behavior for $kL>10$. 

\begin{figure}
\centerline{\includegraphics[width=\figwidth]{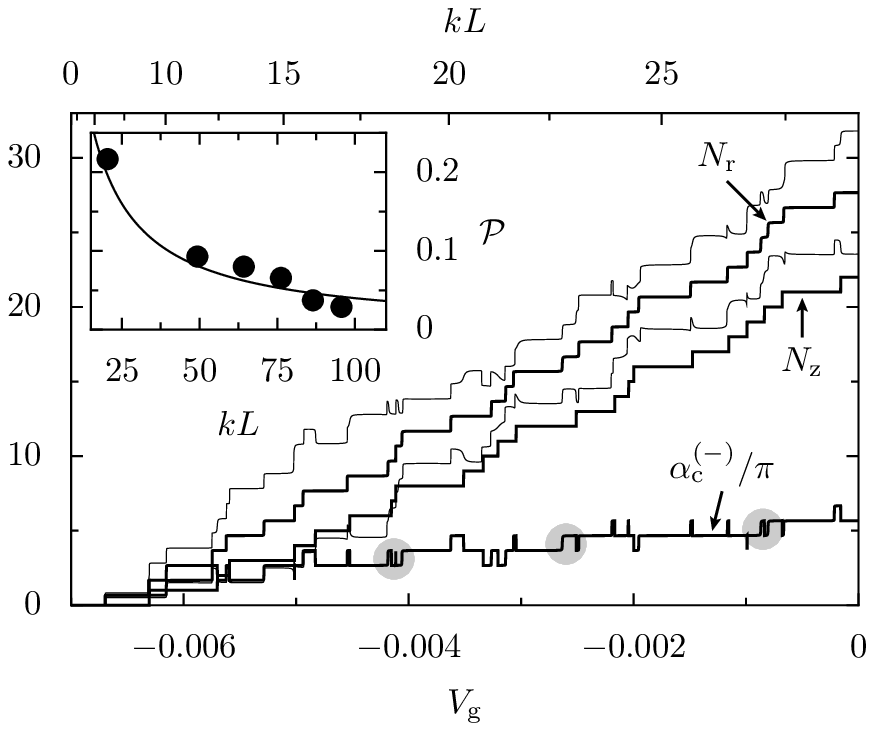}}
\caption{\label{fig:f2} Thick lines: accumulated scattering phase 
$\alpha_{\rm c}^{(-)}/\pi$, and number of resonances (zeros) $N_{\rm r(z)}$ as 
a function of $V_{\rm g}$ (or $kL$). The shaded spots indicate the regions 
where $N_{\rm z}$ lags $N_{\rm r}$. 
They are separated approximately by $\pi$ in $kL$. Thin lines: 
$\alpha_{\rm c}/\pi$ for a small positive and negative magnetic
field. Inset: ${\cal P}=(\Delta N_{\rm r}-\Delta N_{\rm z})/\Delta N_{\rm r}$ 
taken at various intervals of $V_{\rm g}$ as a function of $kL$. The line is a 
guide-to-the-eye that decreases as $(kL)^{-1}$ showing the good agreement of the 
numerical calculations with the prediction of Eq.~(\ref{eq:prob}).}
\label{fig:accpha}
\end{figure}

We next present in Fig.~\ref{fig:accpha} (thick lines) the accumulated scattering 
phase $\alpha_{\rm c}^{(-)}$ (the superscript indicates that we follow the convention used in the experiments of 
taking all phase lapses as $-\pi$), together with the number of accumulated
resonances (zeros) $N_{\rm r(z)}$. $N_{\rm r}$ and $N_{\rm z}$ grow with almost the 
same mean rate for $kL>10$ (at least up to the maximum values of $kL \simeq 100$ that we 
used). In the inset we show the probability
${\cal P}=(\Delta N_{\rm r}-\Delta N_{\rm z})/\Delta N_{\rm r}$
of obtaining more resonances than transmission zeros in an interval of $V_{\rm g}$. 
${\cal P}$ is proportional to the difference of slopes 
$\Delta N_{\rm r}/\Delta V_\mathrm{g}-\Delta N_{\rm z}/\Delta V_\mathrm{g}$.
In agreement with the theoretical analysis, the regions where there are no
zeros between resonances show up with a periodicity (in $kL$) of 
approximately $\pi$ and are separated by intervals where there is an 
alternation of peaks and zeros. 
Upon increasing $k$, sequences exhibiting perfectly alternating resonances and zeros 
become larger and larger, favoring the observation of the universal behavior.

The importance of a magnetic field in the phase-sensitive measurements has
been underlined in experimental \cite{sigrist04} and theoretical \cite{lee99,kim}
works. 
Once we have a magnetic field breaking time-reversal symmetry, the
wave-functions are no longer real, $D_n$ becomes complex, and our analysis in 
terms of the parity of the resonances is no longer applicable. 
We included a magnetic field in our numerical calculations, using a
desymmetrized structure in order to truly break time-reversal symmetry 
\cite{berry}. We observed
that the exact transmission zeros obtained at $B=0$ become minima 
with small (but finite) values. Such a behavior is expected, since a mono-channel 
QD has divergent probability of exhibiting a vanishing $|t|$ when $B=0$, while the 
transmission distribution is uniform in the interval $(0,1)$ for the case where 
time-reversal symmetry is completely broken \cite{JP,jala92,alhassid}.

Avoiding transmission zeros in the complex plane eliminates the $\pi$ phase lapses in favor of 
continuous jumps with large (but finite) derivatives and magnitude $\lesssim\pi$.
At finite fields the ambiguity in the definition of the accumulated phase 
$\alpha_\mathrm{c}$ is then lifted. In Fig.~\ref{fig:accpha} we present 
$\alpha_\mathrm{c}$ for a very small positive (negative) field (thin lines). 
The phase jumps for small positive fields are opposite to those for negative 
fields and the difference in the accumulated phases is of statistical nature
(of the order of the square root of the number of avoided zeros), 
illustrating the diffusion away from the origin in the complex plane for
small $B$ fields. On the other hand, the mean slope of the accumulated phase
is the same in both cases, and coincides with the slope of $\pi N_\mathrm{r}$. 
We point out that the difference between the Friedel and transmission phases 
\cite{taniguchi99,lee99} arises from the ambiguity in the definition of 
$\alpha_\mathrm{c}$ at $B=0$. This ambiguity is lifted if we define 
$\alpha_c(B=0) = {\rm lim}_{B \rightarrow 0^+} \alpha_c(B)$.

Experimentally, the visibility threshold for the AB oscillations can make the
continuous phase evolution at small fields indistinguishable
from the phase lapses at $B=0$. Breaking time-reversal symmetry reduces the 
probability of obtaining very small transmission values and thereby favors the 
observation of a continuous phase evolution as a function of $V_\mathrm{g}$. 
The magnetic field needed for breaking time-reversal symmetry in the QD 
scales as $k^{-1/2}$ \cite{jala92,alhassid}. Experimentally, larger fields have been used
in Ref.~\cite{jacoby95}, where however abrupt phase lapses $\pi$ are enforced by the
use of a two-terminal setup. Refs.~\cite{schuster97,avinum05} on the other hand
used fields too weak to break time-reversal symmetry. 
Ref.~\cite{sigrist04} reported phase lapses, and thus universal behavior, 
in specific magnetic field ranges only, where, presumably, the transmission drops
below the experimental visibility threshold and is thus indistinguishable from a
true zero. We predict that, under the experimental conditions of 
Refs.~\cite{schuster97,avinum05} a magnetic field of the order of 500 $G$ will eliminate 
some of the phase lapses observed at weak fields.

Another test would be to investigate diffusive QD. In a two-dimensional system with 
short-range disorder characterized by an elastic mean-free-path $\ell \ll L$, the right-hand 
side of Eq.~(\ref{BerryBessel}) is suppressed by an exponentially small prefactor 
$e^{-L/2 \ell}$. When there is a sufficient amount of disorder, one obtains ${\cal P}=1/2$, 
as in Ref.~\cite{levy00}, and thus the universal regime disappears. The experiments
report values $\ell \simeq 5 - 15 \mu m$~\cite{jacoby95,schuster97,sigrist04,avinum05}
with QD's of submicron sizes, well in the ballistic regime. We predict that 
a larger, more disordered QD would exhibit a mesoscopic regime with phase lapses
randomly distributed between consecutive resonances with a probability
${\cal P}=1/2$. We also note that the weaker screening of impurities
in few-electron QD makes the dots less 
ballistic, favoring the observation of the mesoscopic regime. 

Up to now we have ignored the effects of the electronic spin. The first such effect is 
a trivial factor of 2 in the density of states that affects the relation between $k$ and $N$.
Second, within the CIM the filling of spin-degenerate one-particle states is by 
pairs \cite{alhassid}. We note that this spin degeneracy would divide by two the 
probability of obtaining, in our analytical and numerical approaches, consecutive 
resonances without a phase lapse, and hence favor the universal behavior. 
Correlation effects may not be negligible in Coulomb blockade valleys, but they are not 
expected to significantly alter the transmission phases close to the resonances nor the 
occurrence of phase lapses in between. 

Our single-particle theory for ballistic QD thus predicts the emergence of a
universal behavior of scattering phases at large $kL$. Our numerics indicate that 
this crossover occurs around $kL \simeq 10$ which, for two-dimensional structures, 
correspond to putting $\sim 15$ electrons on the dot. This is qualitatively in agreement 
with the crossover reported in Ref.~\cite{avinum05} and with the universal behavior 
reported in Refs.~\cite{jacoby95,schuster97}, which work in the range $kL \gtrsim 50$.
We note that another condition is that $kL$ is large enough that the wavefunctions resolve 
the chaoticity of the cavity, which also usually occurs around $kL \simeq 10$. 
We predict the disappearance of the observed universal regime (i) in the presence of a 
large magnetic field and (ii) in larger, more disordered dots in the diffusive regime. 
These predictions could be the basis for a comparison with alternative theories, like
the one of Ref.~[\onlinecite{vonDelft07}].

In conclusion we have provided quantitative, checkable
predictions for the probability of observing long 
sequences of alternating transmission zeros and resonances in scattering through
quantum dots, which
are consistent with the experiments of Refs.~\cite{jacoby95,schuster97,sigrist04,avinum05}.
We stress the probabilistic character of our findings, and
in particular that the absence of phase lapses between resonances is always possible.
We hope that this will stimulate
new experimental investigations.

We thank P.\ Schmitteckert for useful 
discussions. We acknowledge support from the
Spanish MICINN through project FIS2009-07277, the NSF under
grant No DMR-0706319, and the ANR through grant ANR-08-BLAN-0030-02.


\begin{thebibliography}{99}

\bibitem{imry}Y. Imry, {\em Introduction to Mesoscopic Systems}, 2nd ed. 
(Oxford University Press, Oxford, 2002).

\bibitem{jacoby95}
A.\ Yacoby  {\em et al.},
Phys.\ Rev.\ Lett.\ \textbf{74}, 4047 (1995).

\bibitem{schuster97}
R.\ Schuster  {\em et al.}, Nature \textbf{385}, 417 (1997).

\bibitem{sigrist04}M. Sigrist {\em et al.}, Phys.\ Rev.\ Lett \textbf{93}, 
066802 (2004). 

\bibitem{avinum05}
M.\ Avinum-Kalish  {\em et al.},
Nature \textbf{436}, 529 (2005).

\bibitem{levy95}
A.\ Levy Yeyati and M.\ B\"uttiker, Phys.\ Rev.\ B \textbf{52}, R14360 (1995).

\bibitem{aharony2002}A. Aharony  {\em et al.},
Phys.\ Rev.\ B \textbf{66}, 115311 (2002).

\bibitem{lee99}
H.-W.\ Lee, Phys.\ Rev.\ Lett.\ \textbf{82}, 2358 (1999).

\bibitem{taniguchi99}
T.\ Taniguchi and M.\ B\"uttiker, Phys.\ Rev. B \textbf{60}, 13814 (1999).

\bibitem{levy00}
A.\ Levy Yeyati and M.\ B\"uttiker, Phys.\ Rev.\ B \textbf{62}, 7307 (2000).

\bibitem{molina_unp} 
R.A.\ Molina \textit{et al.}, arXiv:1202.1253 [J. Phys. Conf. Ser. (to be published)]. 

\bibitem{molina03} 
R.A.\ Molina  {\em et al.}, Phys.\ Rev.\ B \textbf{67}, 235306 (2003).

\bibitem{alhassid}
Y.\ Alhassid, Rev.\ Mod.\ Phys.\ \textbf{72}, 895 (2000).

\bibitem{JP} 
R.A.\ Jalabert, J.-L.\ Pichard, C.W.J.\ Beenakker, Europhys.\ Lett.\ {\bf 27}, 255 
(1994); H.A.\ Baranger and P.\ Mello, Phys.\ Rev.\ Lett.\ {\bf 73}, 142 (1994);
R.A.\ Jalabert and J.-L.\ Pichard, J.\ Phys.\ (France) {\bf 5}, 287 (1995).

\bibitem{hacken}
G.\ Hackenbroich, Phys.\ Rep.\ \textbf{343}, 463 (2001).

\bibitem{gefen99}
R.\ Baltin and Y.\ Gefen,  Phys.\ Rev.\ Lett.\ \textbf{83}, 5094 (1999). 

\bibitem{silvestrov00} 
P.G.\ Silvestrov and Y.\ Imry, Phys.\ Rev.\ Lett.\ \textbf{85}, 
2565 (2000). 

\bibitem{golosov06}
D.I.\ Golosov and Y.\ Gefen,
Phys.\ Rev.\ B \textbf{74}, 205316 (2006).

\bibitem{bertoni07}
A.\ Bertoni and G.\ Goldoni,
Phys.\ Rev.\ B \textbf{75}, 235318 (2007).

\bibitem{vonDelft07}
C.\ Karrasch \textit{et al.}, Phys.\ Rev.\ Lett.\ \textbf{98}, 186802 (2007). 

\bibitem{goldstein07} M. Goldstein et al., Phys. Rev. B {\bf 79}, 
125307 (2009).

\bibitem{bergfield10} 
J.P.\ Bergfield, Ph.\ Jacquod, and C.A.\ Stafford, Phys.\ Rev.\ B {\bf 82},
205405 (2010).

\bibitem{jala92}
R.A.\ Jalabert, A.D.\ Stone, and Y.\ Alhassid, Phys.\ Rev.\ Lett 
\textbf{68}, 3468 (1992). 


\bibitem{berry77}
M.V.\ Berry, J.\ Phys.\ A \textbf{10}, 2083 (1977). 

\bibitem{urbina04}
J.D.\ Urbina and K.\ Richter, Phys.\ Rev.\ E \textbf{70}, 015201(R) (2004).

\bibitem{narimanov}
E.E.\ Narimanov  {\em et al.}, Phys.\ Rev.\ B \textbf{64}, 
235329 (2001).

\bibitem{kim}
T.-S.\ Kim  {\em et al.}, Phys.\ Rev.\ B \textbf{65}, 
245307 (2002).

\bibitem{berry}
M. Robnik and M.V.\ Berry, J. Phys. A: Math. Gen. {\bf 18}, 1361 (1985).

\end{thebibliography}
\end{document}